\documentclass[11pt]{article}
\pdfoutput=1
\usepackage{amsmath}
\usepackage{amssymb}
\usepackage{graphicx}
\usepackage{hyperref}
\hypersetup{colorlinks,linkcolor=blue,urlcolor=blue}
\DeclareMathOperator{\sign}{sign}
\def\re#1{(\ref{#1})}

\begin{document}
\title{\bf Costly Trading}
\author{Michael Isichenko\footnote{{\tt michael.isichenko@gmail.com}}}
\date{}
\maketitle

\begin{abstract}
We revisit optimal execution of an active portfolio in the presence of
slippage (aka linear, proportional, or absolute-value) costs.  Market
efficiency implies a close balance between active alphas and trading
costs, so even small changes to trading optimization can make a big
difference.  It has been observed for some time that optimal trading
involves a pattern of a no-trade zone with width $\Delta$ increasing
with the slippage cost $c$.  In a setting of a reasonably stable
(non-stochastic) forecast of future returns and a quadratic risk
aversion, it is shown that $\Delta\sim c^{1/2}$, which differs from
the $\Delta\sim c^{1/3}$ scaling reported for stochastic settings.
Analysis of optimal trading employs maximization of a utility
including projected alpha-based profits, slippage costs, and risk
aversion and borrows from a physical analogy of forced motion in the
presence of dry friction.
\end{abstract}

\section{Introduction.  Trading costs and market efficiency}

Trading in financial markets involves costs with security-dependent
structure.  One normally distinguishes between trading costs
proportional to the absolute value of the trade size and those
increasing nonlinearly with the trade size.  The former type includes
fees, taxes, and bid-ask spread and is here referred to as {\em
  slippage}.  The latter, {\em impact} cost is due to a sizable trade
affecting security price.  Empirical measurement of the trading costs
is an important part of active portfolio management and is generally
complicated by the market noise and the difficulty of attribution of
price changes to specific trades.  It generally takes a large database
of orders and executions to run a statistically meaningful trading
cost analysis \cite{BBDJ2018}.

Given the scale of alpha-based portfolios deployed in financial
markets, trading costs are not only significant in quantitative terms
but also bear on qualitative concepts including market efficiency.
The venerable teaching of efficient markets \cite{Samuelson1965} has
been under renewed pressure, at least at the aggregate equity market
level \cite{GK2020,Bouchaud2021}.  At a micro, or security-specific
level, efficiency would mean a poor predictability of future returns
such that, for a regular investor, timing the market is never worth
the risk.  Empirical observations, however, indicate the existence of
thousands of hedge funds, many of them systematic quants ostensibly
knowing what they are doing.  By some estimates \cite{Philippon2015},
the cost of trading in the US markets, defined as profits and wages in
the financial industry, can be as high as 9\% of the GDP.

The persistent presence of actively managed portfolios means that markets
are inefficient at the micro level in the sense that skilled traders,
discretionary or quantitative, are able to generate {\em alphas}, or
reasonably performing forecasts of future returns for specific
securities.  The definition of market efficiency should then be
adjusted to account for the presence of alphas and competing
alpha-based traders.  {\em An efficient market can be defined as one
  in which tradable predictability of security prices is balanced by
  trading costs.}  Such a balance is typical for mid-frequency
statistical arbitrage where the forecasts and the costs are of the
same order of magnitude \cite{Isichenko2021}.  It is then the subtle
difference between the forecast performance\footnote{Performance seems
a better term than accuracy: it is hard to call {\em accurate}
forecasts whose typical correlation with realized returns is of order
1\%.} and the cost of trading, which separates success from failure.
Not surprisingly, the most successful quantitative funds appear to
have a good understanding of their trading costs \cite{Zuckerman2019}
and, by extension, a fairly optimal portfolio construction accounting
for these costs.

The subject of this paper is optimal trading with slippage, which is
the dominant cost for small- to medium-size active equity portfolios.  The
problem was considered in a number of papers starting with
Constantinides \cite{Constantinides1986}, who showed that slippage must
drastically reduce the frequency and volume of optimal trades.
The resulting trading pattern involves a no-trade zone (NTZ) in terms
of position for each security, so it is provably detrimental to trade
while within the NTZ.

The case of slippage is different from impact costs.  The latter
disappear in the limit of a vanishing trading rate and therefore do
not impose a no-trade zone \cite{GP2013,Isichenko2021}.

Formal optimization in the presence of slippage involves some
mathematical challenges due to an absolute-value-type nonlinearity
present in the problem.  Analytical solutions were developed for
stochastic models with security price following a random walk with a
mean-reverting-type drift \cite{MS2011,LDPB2012,MMN2013}.  Stochastic
treatment of slippage optimization uses mathematical methods such as
It\^o algebra or Bellman's optimal control theory. The results of
these analyses indicate that the width of the NTZ is proportional to
the cubic root of slippage, meaning a strong effect of even a small
cost on optimal execution.  An interesting approach based on
heuristics of tracking a cost-free portfolio, but leading to a
computationally expensive Hidden Markov Model, was proposed in
\cite{KR2015}.

The approach adopted in this paper is non-stochastic.  The nonlinear
problem of optimal continuous-time trading with slippage (but no price
impact) and with quadratic risk penalties is solved in a closed form
for a fairly general profile of predicted future returns.  The
solution depends on the convexity properties of the forecast profile.
The scaling for the NTZ width is found to be a square root of
slippage, which is different from the results of stochastic models.

Given a proprietary nature of algorithmic trading, there are
definitely a number of unpublished developments with similar or more
general results.  Eq.~\re{ntz}, one of the main results of this paper,
was derived by the author around 2004; a publication seems warranted
given that the problem has since been discussed in the literature in
considerable detail.

\section{Optimal trading with slippage}

Consider a single security tradable in continuous time so its position
$P(t)$ can take on any real value.  Given a forecast $f(t)$, the
expectation of future security return from current time zero to time
$t$, we can plan the position path $P(t)$ by maximizing a mean-variance
utility functional
\begin{equation}
  U[P(t)] = \int_0^\infty\left[\dot f P - C(\dot P) - kP^2\right]\,dt.
  \label{U}
\end{equation}
Here $C(\dot P)$ is a trading cost rate, normally an even function of
the trading rate $\dot P$, and $k$ is a risk aversion coefficient
needed to penalize for exposure as well as for regularization and
portfolio size control.  It is common to apply a discounting to future
pnl and risk, e.g., by introducing an exponential decay factor in the
integrand of Eq.~\re{U}.  Doing so would have no significant effect on
the discussion below.

In the absence of costs, the optimal position path is
\begin{equation}
  P^{(0)}(t)=\frac{\dot f(t)}{2k}.
  \label{cost-free}
\end{equation}
More generally, a variational maximization of \re{U} gives the
following optimality condition
\begin{equation}
  C''(\dot P)\ddot P - 2kP + \dot f(t) = 0.
  \label{optimality}
\end{equation}
The case of a time-local quadratic cost, $C(\dot P)=\mu\dot P^2$
results in an easily solvable linear equation similar to forced motion
of a body subject to a viscous damping.  A more complicated, but
still tractable situation arises when the viscosity is not local in
time due to a finite lifetime of price impact \cite{Isichenko2021}.

The case of time-local absolute-value (slippage) cost function,
\begin{equation}
  C(\dot P)=c|\dot P|,
  \label{slippage}
\end{equation}
is generally more difficult than quadratic cost and, in mechanical
terms, is similar to dry friction rather than viscosity.  Moving
position $P(t)$ from its current state $P_0$ requires a finite force,
a pattern known as static friction, or
\href{https://en.wikipedia.org/wiki/Stiction}{\em stiction}.  This
sticky point is better seen upon integrating the utility \re{U} by
parts:
\begin{equation}
  U[P(t)] = -\int_0^\infty\left[\dot P(f + c\,\sign(\dot P)) + kP^2\right]\,dt.
  \label{U2}
\end{equation}
This equation clearly compares the magnitude of the forecast $f$ with
the cost $c$ and indicates a threshold behavior of optimal trading.
For slippage cost \re{slippage}, the optimality condition \re{optimality},
\begin{equation}
  c\sign(\dot P)\ddot P - 2kP + \dot f(t) = 0,
  \label{optimality2}
\end{equation}
implies that the optimal path $P(t)$ consists of a combination of
plateaus and the cost-free solution \re{cost-free}, possibly with
finite discontinuities (trades) of $P(t)$.  The solution depends on
the forecast profile $f(t)$ and the initial position $P_0$.  The
analysis of cases can be made more intuitive using the mechanical
analogy of stiction.

\begin{figure}[ht]
  \begin{center}
    \includegraphics[width=\textwidth]{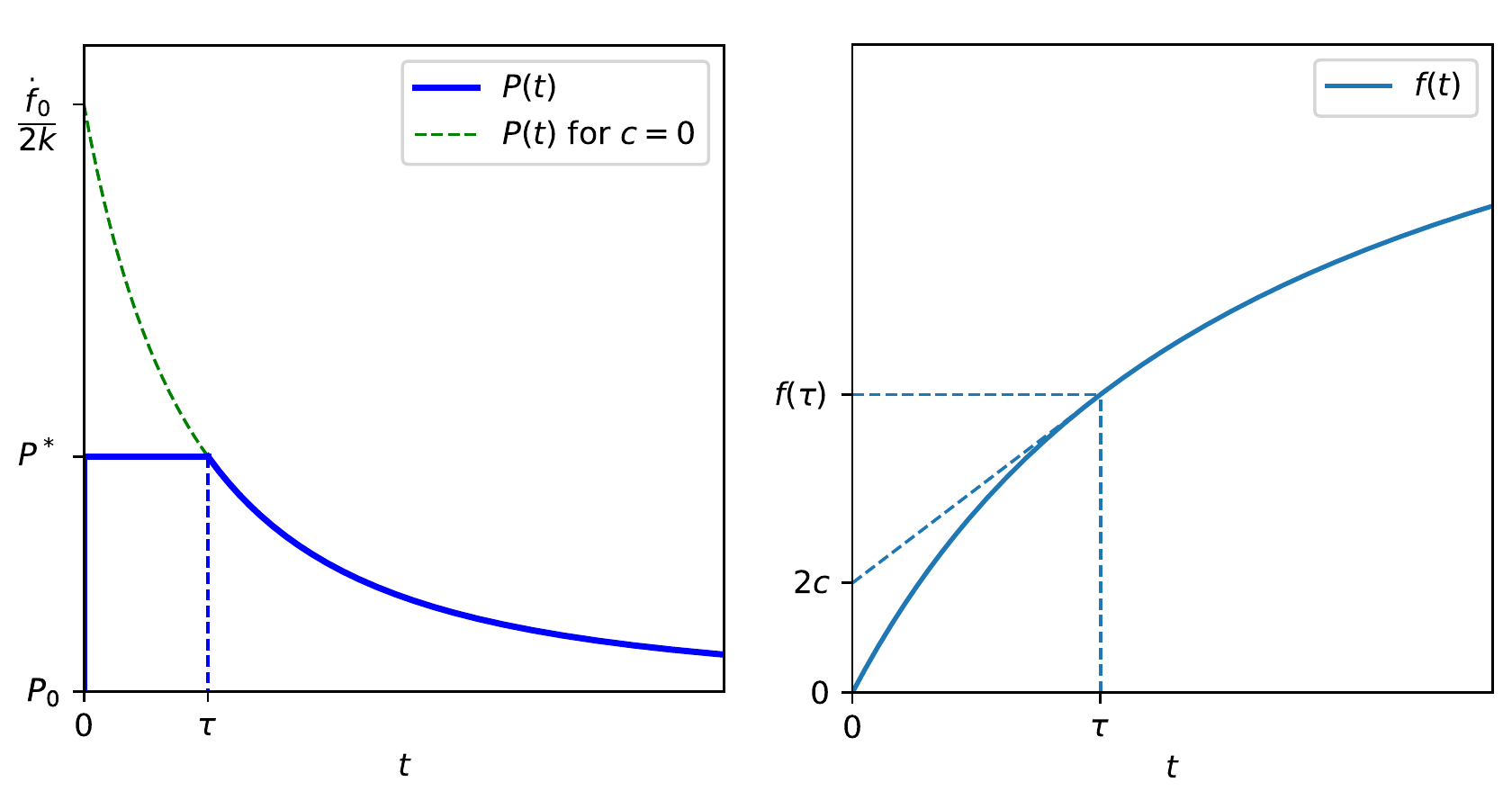}
    \caption{\small Left: optimal position path for a concave forecast
      profile.  Right: geometric solution of optimality
      condition~\re{legendre}.}
    \label{fig:concave}
  \end{center}
\end{figure}

Consider a special, but still fairly generic case of a concave
forecast with $f(0)=0$, $f(\infty)=f_\infty>0$, and $\ddot f(t)<0$
(Fig.~\ref{fig:concave}).
When starting with $P_0=0$, the position is traded to a value $P^*$
smaller than the cost-free target $\dot f(0)/(2k)$ and, upon holding
at that plateau for some time $\tau$, follows \re{cost-free}.  Closing
the position in the long-time limit is forced by risk without reward
as $\dot f(t)\to0$ for $t\to\infty$.  The plateau level $P^*$ is found
by maximizing the utility \re{U} of such path, which now depends on a
single parameter $\tau$:
\begin{equation}
  U(\tau) = \frac{(f(\tau) - 2c)\dot f(\tau)}{2k}
  - \frac{\tau\dot f^2(\tau)}{4k} +\int_\tau^\infty\frac{\dot f^2(t)}{4k}dt.
\end{equation}
The maximum is reached at $U'(\tau)=0$, which condition simplifies to
\begin{equation}
  f(\tau)-\tau\dot f(\tau)=2c.
  \label{legendre}
\end{equation}
Eq.~\re{legendre} gives the plateau duration $\tau$,
\begin{equation}
  \dot f(\tau) = \hat f^{-1}(2c),\quad \hat f(t) \overset{def}{=} \max_\xi(f(\xi)-t\xi),
  \label{legendre2}
\end{equation}
and the initial trade target
\begin{equation}
  P^*=\frac{\dot f(\tau)}{2k}
  =\frac{\hat f^{-1}(2c)}{2k}.
  \label{ntz}
\end{equation}
In Eqs.~\re{legendre2} and \re{ntz}, $\hat f(t)$ is the Legendre
transform (aka convex conjugate) of the forecast $f(t)$ and $\hat
f^{-1}(\cdot)$ means the function inverse. The geometric meaning of
\re{legendre} and \re{legendre2} is that the intercept of a tangent to
the forecast curve $f(t)$ at point $t=\tau$ equals $2c$, the slippage
cost of a roundtrip trade (right Fig.~\ref{fig:concave}).  A finite
solution for $\tau$ in this geometry exists only if the forecast at an
infinite horizon exceeds this cost: $f_\infty>2c$, otherwise such
tangent does not exist and it is best not to trade.

For reasons of forecast revision, as new information becomes
available, the long-term path of position $P(t)$ is less important
than its initial trade (or absence thereof).  A straightforward
analysis of other initial conditions along the same lines as above
shows that the initial optimal trading follows the NTZ pattern: if the
initial position $P_0$ lies between the cost-free target $\dot
f(0)/(2k)$ and $P^*$, it should not be traded.  Otherwise, a trade
should be made to the nearest boundary of the no-trade zone.

The NTZ is not symmetric with respect to the cost-free position
target.  In our treatment, a risk penalty exerts a pressure on the
position toward zero. The cost-free target is then the NTZ boundary
with the larger absolute value; the other boundary is closer to zero
and often equals zero.

The width of the NTZ is data-dependent but its scaling law in the
small-slippage limit must be universal.  A rational function
\begin{equation}
  f(t)=f_\infty\frac{\gamma t}{1 + \gamma t}
\end{equation}
represents a generic smooth forecast profile making our solution
explicit.  For this profile, the lower bound of the NTZ is
\begin{equation}
  P^*=\frac{\gamma}{2k}\left(f_\infty^{1/2} - (2c)^{1/2}\right)^2
  \quad\text{for}\quad f_\infty > 2c.
\end{equation}
For $f_\infty\le 2c$, $P^*=0$.  This example shows that the NTZ width
scales as square root of slippage:
\begin{equation}
  \Delta(c)=\frac{\dot f_0}{2k} - P^*=\frac{\dot f_0}{k}\left(\frac{2c}{f_\infty}\right)^{1/2} + O(c)\quad
    \text{for}\quad c\ll f_\infty.
\end{equation}

It is also necessary to consider a non-concave forecast term structure
$f(t)$, because the forecast is generally a combination of multiple
signals of different horizons.  If $f(t)$ has a single inflection
point, a two-plateau solution can be similarly optimized.  The result
is also expressible via the Legendre transform but the algebra is
uglier.  For more complicated forecast profiles, there is not much
hope for a manageable analytical solution, even with the help of
computer algebra, but the pattern of no-trade zone is universal.

\begin{figure}[ht]
  \begin{center}
    \includegraphics[width=\textwidth]{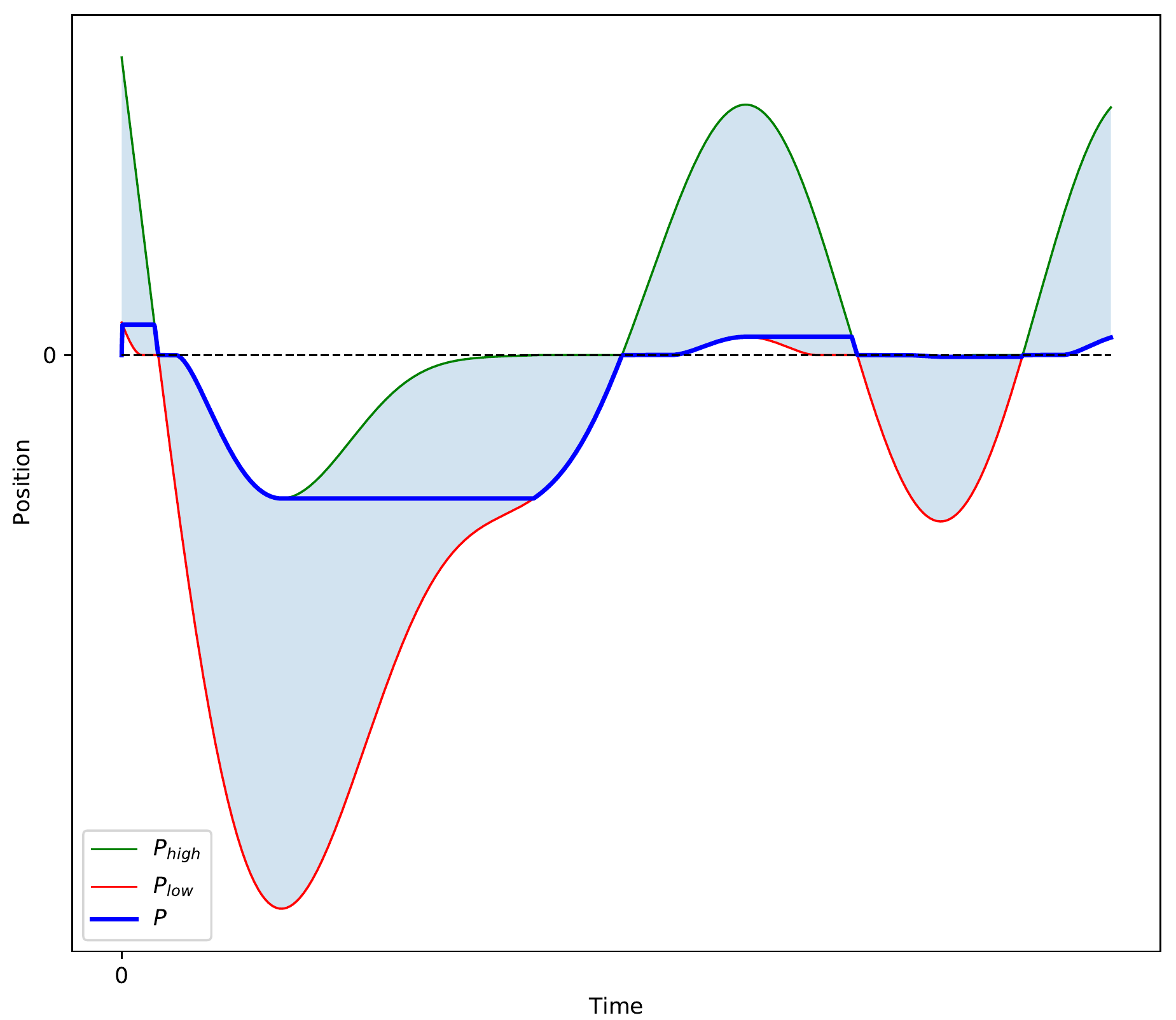}
    \caption{\small Schematic of trading with no-trade zone
      $[P_{low},P_{high}]$, the shaded area.  Optimal security
      position $P(t)$ is driven by the evolving NTZ boundaries.}
    \label{fig:ntz}
  \end{center}
\end{figure}

As discussed below, the process of forecast revision makes the
forecast curve dependent on both the horizon $t$ and the time $t_0$ of
forecasting, which was fixed at zero in the analysis above.  The exact
details of the solution \re{ntz} may not hold exactly, but the the
predicted NTZ pattern will hold and its bounds can be estimated.  An
optimal trading would generally involve the security position $P(t_0)$
forced inside an evolving NTZ, as depicted in Fig.~\ref{fig:ntz}.

\section{Discussion}

The NTZ optimal trading pattern in the presence of slippage costs is
quite intuitive due to a mechanical analogy of forced motion with dry
friction.  The same pattern is present in
\href{https://en.wikipedia.org/wiki/Total_variation_denoising}{total
  variation denoising}  seeking a smoother function approximating
noisy data where the sum of absolute values of the function changes is
penalized \cite{Condat2013}.  Portfolio trading is essentially the same
type of smoothing when chasing noisy alpha opportunities.

A non-stochastic forward-looking position path optimization predicts a
no-trade zone qualitatively similar to the pattern described in
stochastic models, but the scaling of the NTZ width is somewhat
different from the stochastic case.  A common conclusion of several
quantitative analyses is that the NTZ size is a nonlinear function of
trading slippage which doesn't quickly go away as the cost gets
smaller (even if it does).  A fairly general closed-form solution
\re{ntz} is available for a fixed-convexity forecast profile and is
generalizable to other profiles.  A more practical setting would
use a degree of stochasticity to describe the process of forecast
revision.  Accounting for the forecast revision would lead to a wider
NTZ due to a higher expected number of roundtrips.  This heuristic
agrees with the wider NTZ $\Delta\sim c^{1/3}$ in a fully stochastic
setting vs the deterministic scaling $\Delta\sim c^{1/2}$.

Slippage costs are also subject to fluctuations and could benefit from
a stochastic treatment.  A simpler approach is to distinguish between
the mean slippage $c$ applicable to future trades and the current
(initial) slippage $c_0$ which can be predicted with a higher
accuracy.  This amounts to replacing $2c$ in \re{ntz} by $c_0+c$ and
thereby gauging the NTZ depending on current execution opportunities,
e.g., by combining active orders with liquidity provision.

Optimal trading of a portfolio, i.e., multiple securities interacting
via risk factors, also involves no-trade zones due to slippage costs,
because the interaction of securities in a portfolio utility function
acts like a ``risk pressure'' correction to the forecast
\cite{Isichenko2021}.  Even in continuous time, portfolio trading
pattern will generate discrete events of position adjustment.
Qualitatively, one can imagine a particle in an evolving
multi-dimensional potential well.  The particle (a vector of portfolio
positions) is drawn to the bottom of the well but is also subject to a
slippage {\em stiction} which is only occasionally overcome when and
where the slope of the well exceeds a threshold.

The effect of even small slippage costs is significant for an actively
managed portfolio unless a sufficiently strong and low-turnover alpha
is used.  It is fair to assume that most alpha-driven portfolios
underestimate market friction (or its implications for portfolio
construction) and trade too much.

\end{document}